\documentclass[twocolumn]{aastex701}

\begin{document}



\title{Reconstructing the Projected Dark Matter Field across 
\(0.1\)--\(100\,\mathrm{Mpc}\) Scales from the SDSS Survey
\footnote{This is the sixteenth paper in the ``From Halos to Galaxies'' series.}}

\author[gname=Kunyao,sname=Zhao,0009-0000-3647-6527]{Kunyao Zhao}
\email{} 
\affiliation{Department of Astronomy, School of Physics, Peking University, 5 Yiheyuan Road, Beijing 100871, People's Republic of China}
\affiliation{Kavli Institute for Astronomy and Astrophysics, Peking University, 5 Yiheyuan Road, Beijing 100871, People's Republic of China}

\author[0000-0003-0939-9671]{Yingjie Peng}
\email{yjpeng@pku.edu.cn}
\affiliation{Department of Astronomy, School of Physics, Peking University, 5 Yiheyuan Road, Beijing 100871, People's Republic of China}
\affiliation{Kavli Institute for Astronomy and Astrophysics, Peking University, 5 Yiheyuan Road, Beijing 100871, People's Republic of China}

\correspondingauthor{Yingjie Peng}
\email{yjpeng@pku.edu.cn}

\author[gname=Dingyi,sname=Zhao,0009-0001-1564-3944]{Dingyi Zhao}
\email{} 
\affiliation{Department of Astronomy, School of Physics, Peking University, 5 Yiheyuan Road, Beijing 100871, People's Republic of China}
\affiliation{Kavli Institute for Astronomy and Astrophysics, Peking University, 5 Yiheyuan Road, Beijing 100871, People's Republic of China}

\author[gname=Xiaohu,sname=Yang,0000-0003-3997-4606]{Xiaohu Yang}
\email{} 
\affiliation{Tsung-Dao Lee Institute, and Shanghai Key Laboratory for Particle Physics and Cosmology, Shanghai Jiao Tong University, Shanghai 200240, People's Republic of China}
\affiliation{Department of Astronomy, School of Physics and Astronomy, Shanghai Jiao Tong University, Shanghai 200240, People's Republic of China}

\author[gname=Luis C.,sname=Ho,0000-0001-6947-5846]{Luis C. Ho}
\email{} 
\affiliation{Kavli Institute for Astronomy and Astrophysics, Peking University, 5 Yiheyuan Road, Beijing 100871, People's Republic of China}
\affiliation{Department of Astronomy, School of Physics, Peking University, 5 Yiheyuan Road, Beijing 100871, People's Republic of China}

\author[gname=Kai,sname=Wang,0000-0002-3775-0484]{Kai Wang}
\email{} 
\affiliation{Institute for Computational Cosmology, Department of Physics, Durham University, South Road, Durham DH1 3LE, UK}
\affiliation{Centre for Extragalactic Astronomy, Department of Physics, Durham University, South Road, Durham DH1 3LE, UK}

\author[gname=Jing,sname=Dou,0000-0002-6961-6378]{Jing Dou}
\email{} 
\affiliation{National Astronomical Observatories, Chinese Academy of Sciences, Beijing 100101, People's Republic of China}

\author[gname=Zeyu,sname=Gao,0000-0002-0182-1973]{Zeyu Gao}
\email{} 
\affiliation{Department of Astronomy, School of Physics, Peking University, 5 Yiheyuan Road, Beijing 100871, People's Republic of China}
\affiliation{Kavli Institute for Astronomy and Astrophysics, Peking University, 5 Yiheyuan Road, Beijing 100871, People's Republic of China}

\author[gname=Qiusheng,sname=Gu,0000-0002-3890-3729]{Qiusheng Gu}
\email{} 
\affiliation{School of Astronomy and Space Science, Nanjing University, Nanjing 210093, People's Republic of China}

\author[gname=Yukun,sname=Liu,]{Yukun Liu}
\email{} 
\affiliation{Department of Astronomy, School of Physics, Peking University, 5 Yiheyuan Road, Beijing 100871, People's Republic of China}
\affiliation{Kavli Institute for Astronomy and Astrophysics, Peking University, 5 Yiheyuan Road, Beijing 100871, People's Republic of China}

\author[gname=Roberto,sname=Maiolino,0000-0002-4985-3819]{Roberto Maiolino}
\email{} 
\affiliation{Cavendish Laboratory, University of Cambridge, 19 J.J. Thomson Avenue, Cambridge CB3 0HE, UK}
\affiliation{Kavli Institute for Cosmology, University of Cambridge, Madingley Road, Cambridge CB3 0HA, UK}
\affiliation{Department of Physics and Astronomy, University College London, Gower Street, London WC1E 6BT, UK}

\author[gname=Houjun,sname=Mo,0000-0001-5356-2419]{Houjun Mo}
\email{} 
\affiliation{Department of Astronomy, University of Massachusetts, Amherst, MA 01003, USA}

\author[gname=Alvio,sname=Renzini,0000-0002-7093-7355]{Alvio Renzini}
\email{} 
\affiliation{INAF--Osservatorio Astronomico di Padova, Vicolo dell'Osservatorio 5, I-35122 Padova, Italy}

\author[gname=Canpo,sname=Su,]{Canpo Su}
\email{} 
\affiliation{Department of Astronomy, School of Physics, Peking University, 5 Yiheyuan Road, Beijing 100871, People's Republic of China}
\affiliation{Kavli Institute for Astronomy and Astrophysics, Peking University, 5 Yiheyuan Road, Beijing 100871, People's Republic of China}

\author[gname=Bitao,sname=Wang,0000-0002-6137-6007]{Bitao Wang}
\email{} 
\affiliation{School of Physics and Electronics, Hunan University, Changsha 410082, People's Republic of China}

\author[gname=Yu-Chen,sname=Wang,0000-0002-8429-7088]{Yu-Chen Wang}
\email{} 
\affiliation{Department of Astronomy, School of Physics, Peking University, 5 Yiheyuan Road, Beijing 100871, People's Republic of China}
\affiliation{Kavli Institute for Astronomy and Astrophysics, Peking University, 5 Yiheyuan Road, Beijing 100871, People's Republic of China}

\author[gname=Bingxiao,sname=Xu,]{Bingxiao Xu}
\email{} 
\affiliation{Kavli Institute for Astronomy and Astrophysics, Peking University, 5 Yiheyuan Road, Beijing 100871, People's Republic of China}

\author[gname=Feng,sname=Yuan,0000-0003-3564-6437]{Feng Yuan}
\email{} 
\affiliation{Center for Astronomy and Astrophysics and Department of Physics, Fudan University, Shanghai 200438, People's Republic of China}

\author[gname=Xingye,sname=Zhu,0000-0002-9529-1044]{Xingye Zhu}
\email{} 
\affiliation{Department of Astronomy, School of Physics, Peking University, 5 Yiheyuan Road, Beijing 100871, People's Republic of China}
\affiliation{Kavli Institute for Astronomy and Astrophysics, Peking University, 5 Yiheyuan Road, Beijing 100871, People's Republic of China}

\begin{abstract}

Dark matter sets the gravitational environment in which galaxies form and evolve, but cannot be observed directly. We present a conditional diffusion model that reconstructs the projected dark matter density field from the galaxy stellar-mass density field for direct application to galaxy surveys. The model is trained on CAMELS and validated on the independent IllustrisTNG300-1 simulation. Halo masses inferred from the reconstructed projected-aperture measurements agree well with the corresponding true values, with a scatter below 0.2 dex. On 100 kpc scales, reconstructed surface densities show a typical scatter of $\sim 0.3$ dex in the regime most relevant for observations. We apply the model to SDSS galaxies with $\mathrm{M_\star \ge 10^9\,M_\odot}$ in a contiguous low-redshift region. Averaging over 100 stochastic realizations, we reconstruct and publicly release a projected dark matter field covering $90\times90\,(h^{-1}\,\mathrm{Mpc})^2$ with a pixel size of $0.097\,h^{-1}\,\mathrm{Mpc}$. This pixel area corresponds to the characteristic projected area of halos with masses of $\sim 10^{10.6}\,h^{-1}\,M_\odot$. The map reveals the multiscale projected cosmic web, including cluster-scale overdensities, filaments and voids. Projected-aperture masses are statistically consistent with SDSS group-catalog masses, while the derived halo mass function broadly matches mock-catalog expectations. The reconstructed projected potential places Coma in one of the deepest wells and near a convergence region of the inferred projected acceleration field, suggesting that the reconstruction retains both local overdensities and coherent large-scale projected gravitational structure. This work shows that diffusion-based dark matter reconstruction can be applied to real galaxy surveys, enabling halo-mass- and spatially resolved dark-matter-environment-based studies of galaxy evolution in SDSS and future wide-area surveys.

\end{abstract}


\section{Introduction} 

Dark matter provides the gravitational backbone of the Universe. In the standard $\Lambda$CDM paradigm, its nonlinear evolution drives the growth of structure, giving rise to dark matter halos, the filamentary cosmic web, and the large-scale environments in which galaxies form and evolve \citep{white_1978,Peeb_1980,blum_1984,Bond_1996}. Characterizing the dark matter distribution is therefore central to both cosmology and galaxy evolution. It is the underlying mass field, rather than the galaxy distribution alone, that sets the gravitational environments governing halo assembly, gas accretion, satellite infall, and many of the processes that shape galaxy properties.

However, dark matter cannot be observed directly. Its distribution is instead inferred from tracers of the underlying matter field. One of the principal observational probes is weak gravitational lensing, which reconstructs projected mass from the coherent distortions of background galaxy shapes \citep{Kais_1993,Tayl_2004,Mass_2010}. In practice, the precision is limited by line-of-sight projection effects, shape noise and related observational systematics \citep{Bart_2001,Jeff_2021}. As a result, obtaining spatially resolved dark matter maps over wide areas and across a broad dynamic range remains challenging.



 Galaxies provide another important tracer of the matter field. Because galaxies reside in dark matter halos, their distribution traces the underlying matter field, albeit in a biased, nonlinear, and environmentally dependent manner \citep{Wech_2018,peng_2020,wang_2025}. A long-standing goal has therefore been to reconstruct the dark matter field from galaxy data as directly as possible. 

Early reconstruction methods included constrained Gaussian realizations \citep{bert_1987,hoff_1991,van_1996}, as well as Wiener-filter estimates of the large-scale density and velocity fields \citep{wien_1995,fish_1995}. Such constraints were subsequently used to construct cosmological initial conditions that could be evolved numerically to generate constrained simulations of the nearby Universe \citep{bist_1998,krav_2002,klyp_2003,gott_2010}. More recent Bayesian forward-modeling approaches infer the initial density field by combining a structure-formation model with a likelihood for the observed galaxy distribution. In particular, \citet{kita_2012} presented one of the first forward-model reconstruction of the local cosmic density and peculiar velocity fields from observed galaxy data, inferring initial conditions consistent with the 2MRS galaxy distribution while accounting for coherent redshift-space distortions. Subsequent studies improved the inference of cosmological initial conditions and the modeling of their nonlinear evolution \citep{jasc_2013,kita_2013,wang_2013,jasc_2019}. Related work then evolved the reconstructed initial conditions in constrained numerical simulations to produce detailed realizations of the nearby Universe \citep{hess_2013,wang_2014,wang_2016}. However, these reconstructions necessarily depend on the adopted cosmology and dynamical model, and their finite effective resolution may limit object-by-object studies of halo-scale environments.

Machine learning (ML) approaches provide a complementary route. Instead of prescribing an explicit analytic model for tracer bias, they learn the mapping between observable tracers and the underlying matter field. Recent work has shown the promise of this strategy. For example, \citet{kryw_2025} compared linear, halo-model, and GNN-CNN methods for recovering dark matter and baryon density fields, showing the strong potential of ML in simulations. Other studies have extended ML-based methods to observational applications, including reconstructions of the local cosmic web and the three-dimensional density fields from nearby galaxy catalogs and DESI-like samples \citep{Hong_2021,Wang_2023,Lilow_2024,shi_2025}. However, most of these methods rely on direct regression. While effective on large scales, they are generally less well suited to capturing the full non-Gaussian, multi-scale spatial distribution of the matter field over wide dynamic range.


Diffusion probabilistic models are particularly attractive in this context. By iteratively denoising a latent field, they can learn complex distributions and recover highly non-Gaussian structure \citep{ho_2020, King_2021}. This makes them well suited for the reconstruction of dark matter, whose distribution is strongly nonlinear on halo and filament scales. A recent example is the diffusion-based method of \citet{ono_2024}, which infers dark matter distribution from the stellar-mass field using the CAMELS simulations. This result is promising, but the model uses the full stellar particle distribution as input, which is not directly available in galaxy surveys. In a follow-up study, \citet{park_2024} extended the diffusion-based method to the Cosmicflows-3 catalog. Yet Cosmicflows-3 is a heterogeneous compilation of local distance measurements rather than a galaxy survey with a well-defined selection function and footprint, and their analysis did not explicitly model redshift-space effects. It remains important to establish whether diffusion-based dark matter reconstruction can be carried through to real galaxy survey data using inputs that are directly observable.


This question has become increasingly timely in the era of modern spectroscopic and imaging surveys, including the Dark Energy Spectroscopic Instrument
\citep[DESI;][]{DESI_2016a,DESI_2016b,desi_2024,desi_2026}, the Vera C. Rubin Observatory Legacy
Survey of Space and Time \citep[LSST;][]{ivez_2019}, and the Chinese Space Station Survey Telescope \citep[CSST;][]{Gong_2019,gong_2025,Zhan_2021,miao_2023,csst_2026}. 
These surveys are delivering an increasingly detailed view of large-scale structure and galaxy environments. Yet most observational studies of galaxy environment still rely primarily on galaxy number density or related overdensity estimators. A practical method for reconstructing the dark matter field from galaxy survey data would enable descriptions based more directly on halo mass and dark matter environment.


In this work, we present a variational diffusion model for reconstructing the projected dark matter field from a projected galaxy stellar-mass density field constructed from galaxies with $M_\star \ge 10^9\,M_\odot$. The input is designed for direct application to survey data. We train the model on the Cosmology and Astrophysics with MachinE Learning Simulations (CAMELS) \citep{camels_2021,Vill_2023,ni_2023}, validate it on the IllustrisTNG300-1 simulation \citep{Springel_2017,Pill_2017,nelson_2021}, and then apply it to a contiguous low-redshift region of the SDSS survey. The resulting projected field covers a $90\times90\,(h^{-1}\,\mathrm{Mpc})^2$ region with a pixel size of $0.097\,h^{-1}\,\mathrm{Mpc}$. At this sampling scale, halos above approximately $10^{10.6}\,h^{-1}\,M_\odot$ have characteristic projected areas larger than one map-pixel area. It captures multiscale features of the projected matter distribution, including cluster-scale overdensities, filamentary connections, and extended underdense regions. In addition, the consistency of the projected-aperture mass estimates with the SDSS group catalog, the broad agreement of the inferred mass function with the L-GALAXIES mock catalog, and the coherent patterns in the projected potential field indicate that the reconstruction retains meaningful large-scale information beyond local density peaks.

This work shows that a conditional diffusion model can be applied to real SDSS observations for projected dark matter reconstruction. A key feature of the present method is its focus on statistically characterizing projected halo-scale structures around individual galaxy groups, enabling galaxy properties to be studied in relation to their surrounding matter environments. It therefore provides a framework for moving observational studies of galaxy environment from traditional galaxy-overdensity-based descriptions toward halo-mass- and dark-matter-environment-based ones.

The paper is organized as follows. In Section~\ref{sec:data}, we describe the simulation and observational datasets used in this work. In Section~\ref{sec:model}, we present the diffusion model and training strategy. Section~\ref{sec:results} shows the main results, including validation on an independent simulation and application to the SDSS galaxy survey. Finally, Section~\ref{sec:summary} summarizes the main conclusions and discusses future directions. In this paper, we adopt a Planck cosmology \citep{planck_2014} with $\sigma_8 = 0.829$, $H_0 = 67.3\ \mathrm{km\ s^{-1}\ Mpc^{-1}}$, $\Omega_\Lambda = 0.685$, $\Omega_{\rm m} = 0.315$, $\Omega_{\rm b} = 0.04857$, and $n_{\rm s} = 0.96$. All spatial coordinates and map scales are given in comoving units.


\begin{figure*}[t]
\centering
\includegraphics[width=\textwidth]{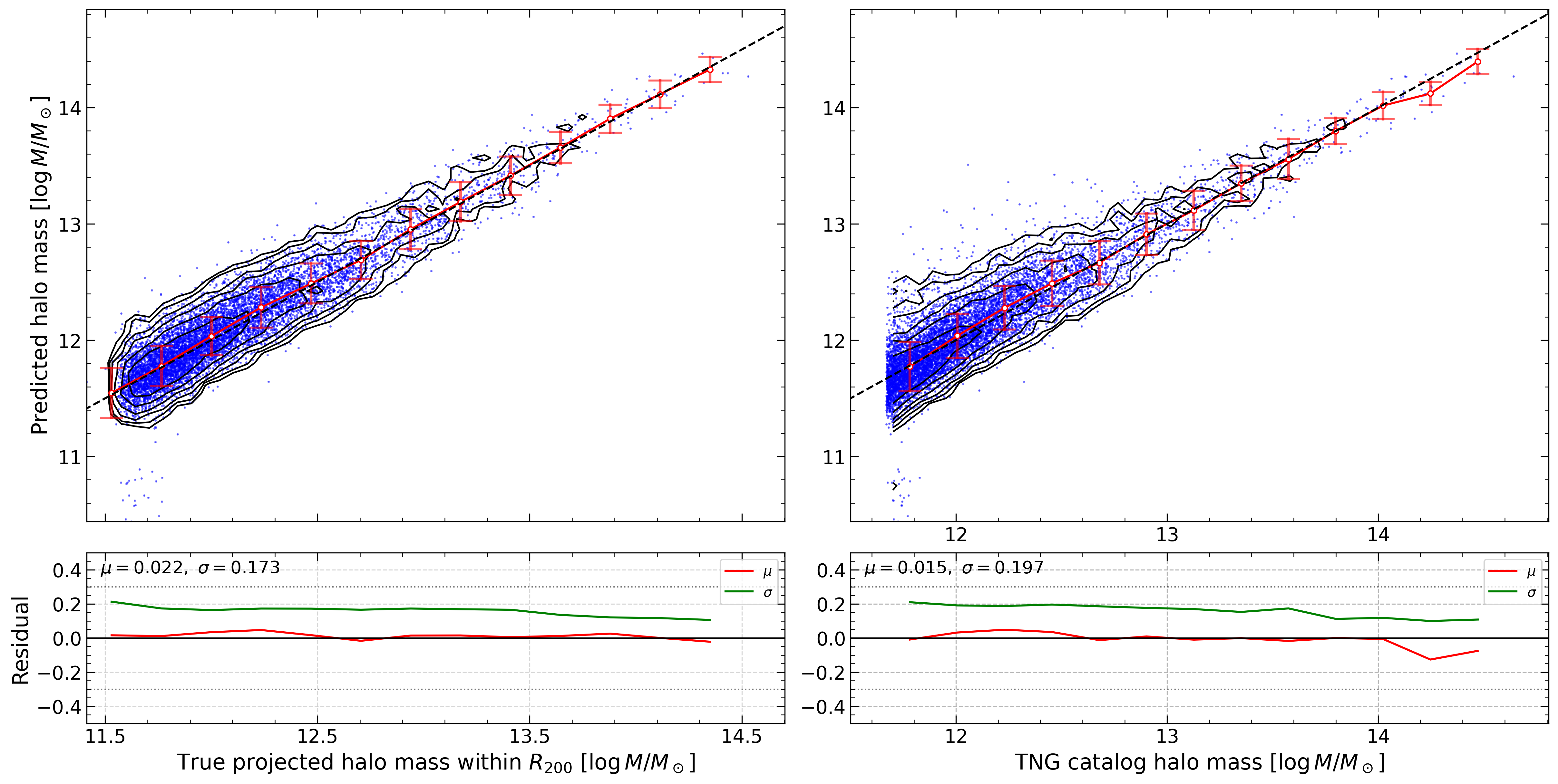}
\caption{\textbf{ Halo-level validation of the reconstruction in the IllustrisTNG300-1 simulation.} \textbf{Left panel:} Comparison between the projected masses integrated within $\mathrm{R_{200}}$ in the ground-truth and reconstructed maps. Blue points denote individual dark matter halos, and black contours represent the number density of these halos. The red line shows the mean predicted mass within each true halo mass bin; error bars indicate the standard deviation. The black dashed line denotes the one-to-one relation between true and predicted halo mass. In the inset, the red and green lines indicate the mean ($\mu$) and standard deviation ($\sigma$) of the residuals within each true halo mass bin respectively, with the gray dashed lines representing $\pm \,0.3$ dex. \textbf{Right panel:} Comparison between predicted projected mass within $\mathrm{R_{200}}$ in the reconstructed map and $\mathrm{M_{200}}$ of groups in the IllustrisTNG300-1 catalog. Visual conventions follow those used in the left panel. All halo masses are given in units of $M_\odot$ and shown on a logarithmic scale. 
}\label{fig1:tng}
\end{figure*}


\begin{figure}[t]
\centering
\includegraphics[width=\columnwidth]{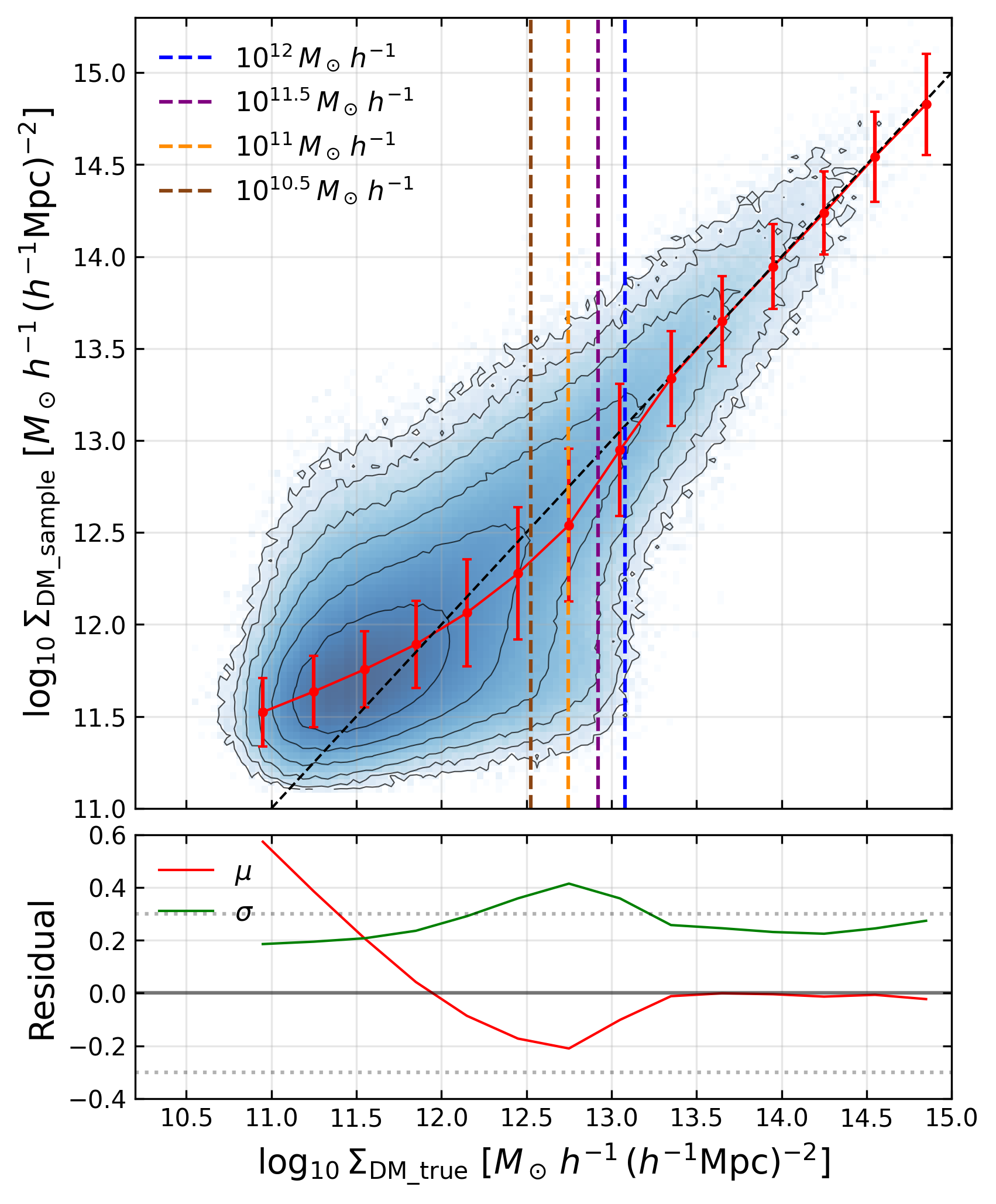}
\caption{\textbf{Pixel-level comparison between the true and reconstructed projected dark matter surface density field in IllustrisTNG300-1.} Black contours and the blue shading represent the logarithmic number density distribution based on $2100 \times 2100$ pixels, and the black dashed line marks the one-to-one relation for reference. The red line gives the mean value within each true density bin, with error bars showing the standard deviation. The four colored vertical dashed lines mark the characteristic surface densities of the least massive halos above the mass thresholds $10^{10.5}\,M_{\odot}\,h^{-1}$, $10^{11}\,M_{\odot}\,h^{-1}$, $10^{11.5}\,M_{\odot}\,h^{-1}$, and $10^{12}\,M_{\odot}\,h^{-1}$. The lower panel shows the mean (red line) and standard deviation (green line) of the residuals between the reconstructed and true values, while the gray dashed lines indicate $\pm 0.3$. All quantities are expressed in logarithmic surface density units of 
$M_\odot\,h^{-1}(h^{-1}\,\mathrm{Mpc})^{-2}$. }
\label{fig5:pixel}
\end{figure}


\begin{figure*}[t]
\centering
\includegraphics[width=\textwidth]{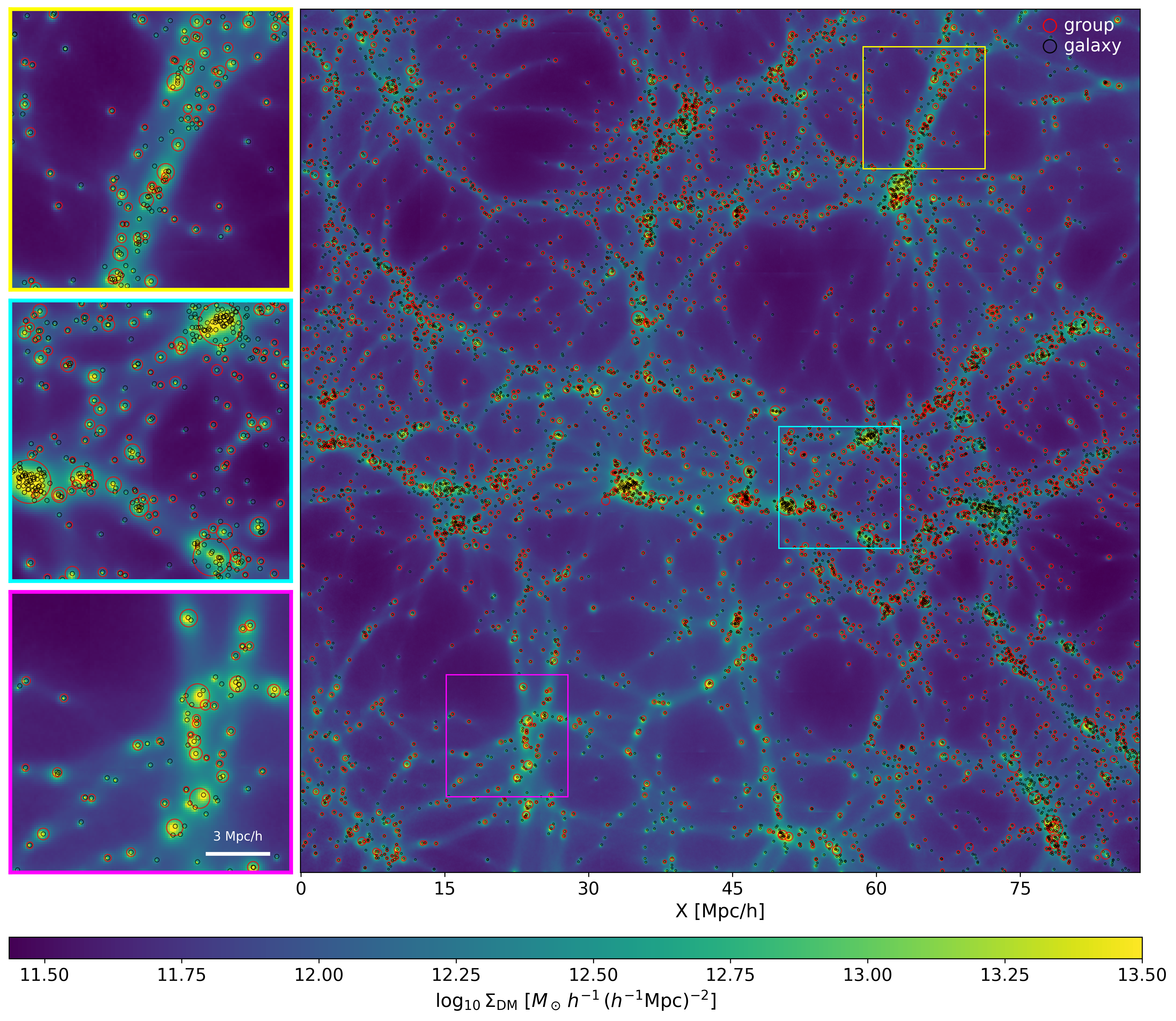} \caption{ \textbf{\boldmath Reconstructed projected dark matter density field over a $90 \times 90~(h^{-1}\mathrm{Mpc})^2$ region from the SDSS survey at $z \sim 0.03$.} The map has a pixel size of $0.097\,h^{-1}\,\mathrm{Mpc}$ and is obtained by averaging 100 stochastic realizations generated from different initial random seeds. The colour scale encodes the logarithmic surface mass density, shown in units of $M_\odot\,h^{-1}(h^{-1}\,\mathrm{Mpc})^{-2}$. Black circles mark the positions of the input galaxies with $\mathrm{M_\star \geq 10^9~M_\odot}$, and red circles indicate galaxy groups from the SDSS group catalog \citep{Yang_2005,Yang_2007,Zhao_2025}, with circle sizes proportional to their radii. Three zoom-in panels on the left provide detailed views of selected subregions, corresponding to the coloured boxes on the right side. Each panel spans approximately $14\,h^{-1}\,\mathrm{Mpc}$ and a $3\,h^{-1}\,\mathrm{Mpc}$ scale bar is provided for reference.
}\label{fig2:dmimage}
\end{figure*}

\begin{figure*}[t]
\centering
\includegraphics[width=\textwidth]{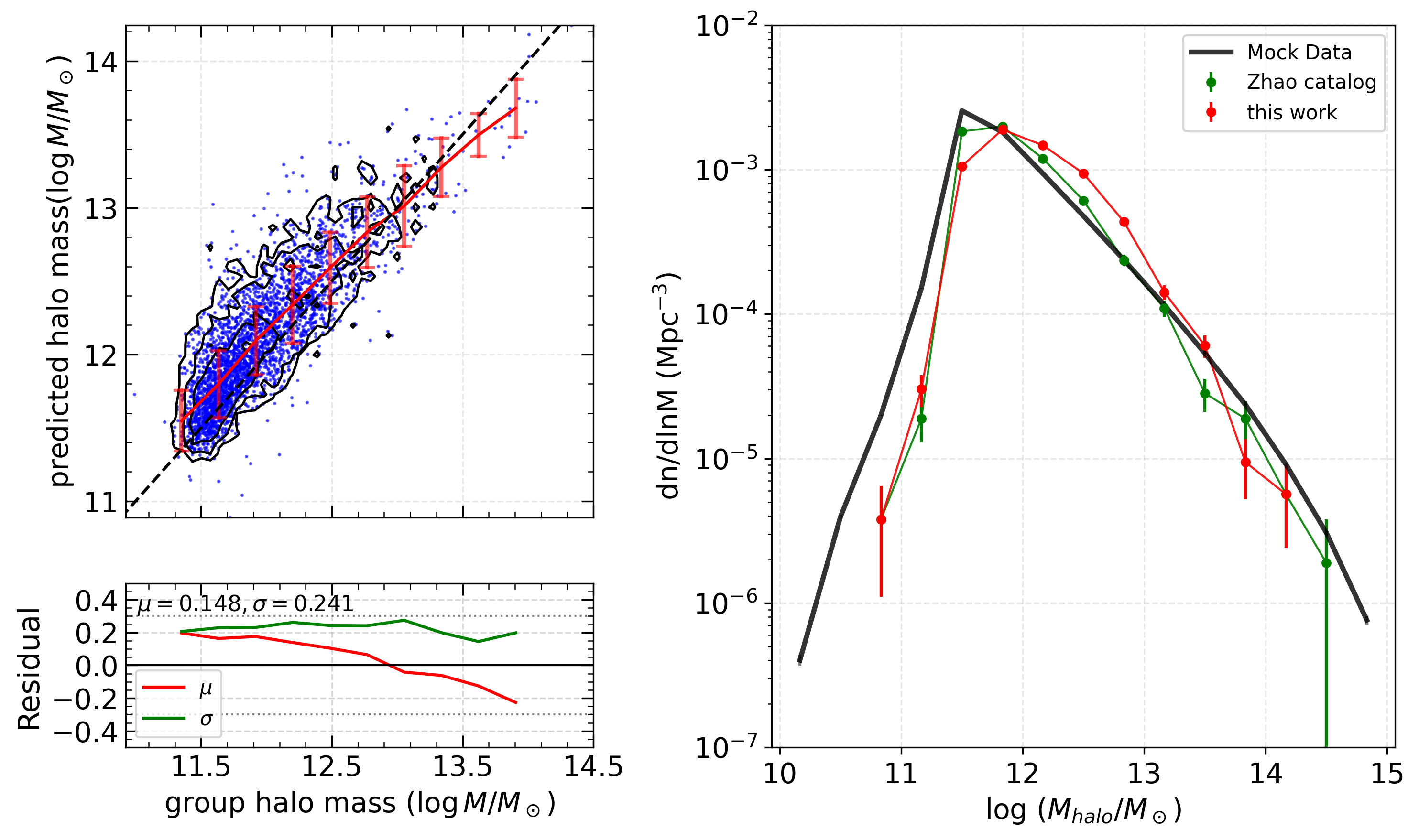} \caption{\textbf{ Halo-scale mass comparison and halo mass functions in the SDSS reconstruction.} \textbf{Left panel:} Comparison between the reconstruction-based projected masses in this work and those listed in the Zhao group catalog. Blue points denote individual halos, while the black contours indicate the number density distribution. The red line represents the mean predicted mass in each halo mass bin from the Zhao catalog, and the vertical error bars indicate the standard deviation. The lower left inset shows the mean (red line) and standard deviation (green line) of the residuals for each halo mass bin, with the gray dashed lines indicating $\pm \,0.3$ dex. \textbf{Right panel:} Halo mass functions derived from the reconstruction (red), the Zhao catalog (green), and the L-GALAXIES mock catalog (black) over the same redshift range. The error bars represent the statistical uncertainty in the number of halos within each halo mass bin.}\label{fig3:halo} 
\end{figure*}

\begin{figure*}[t]
\centering
\includegraphics[width=\textwidth]{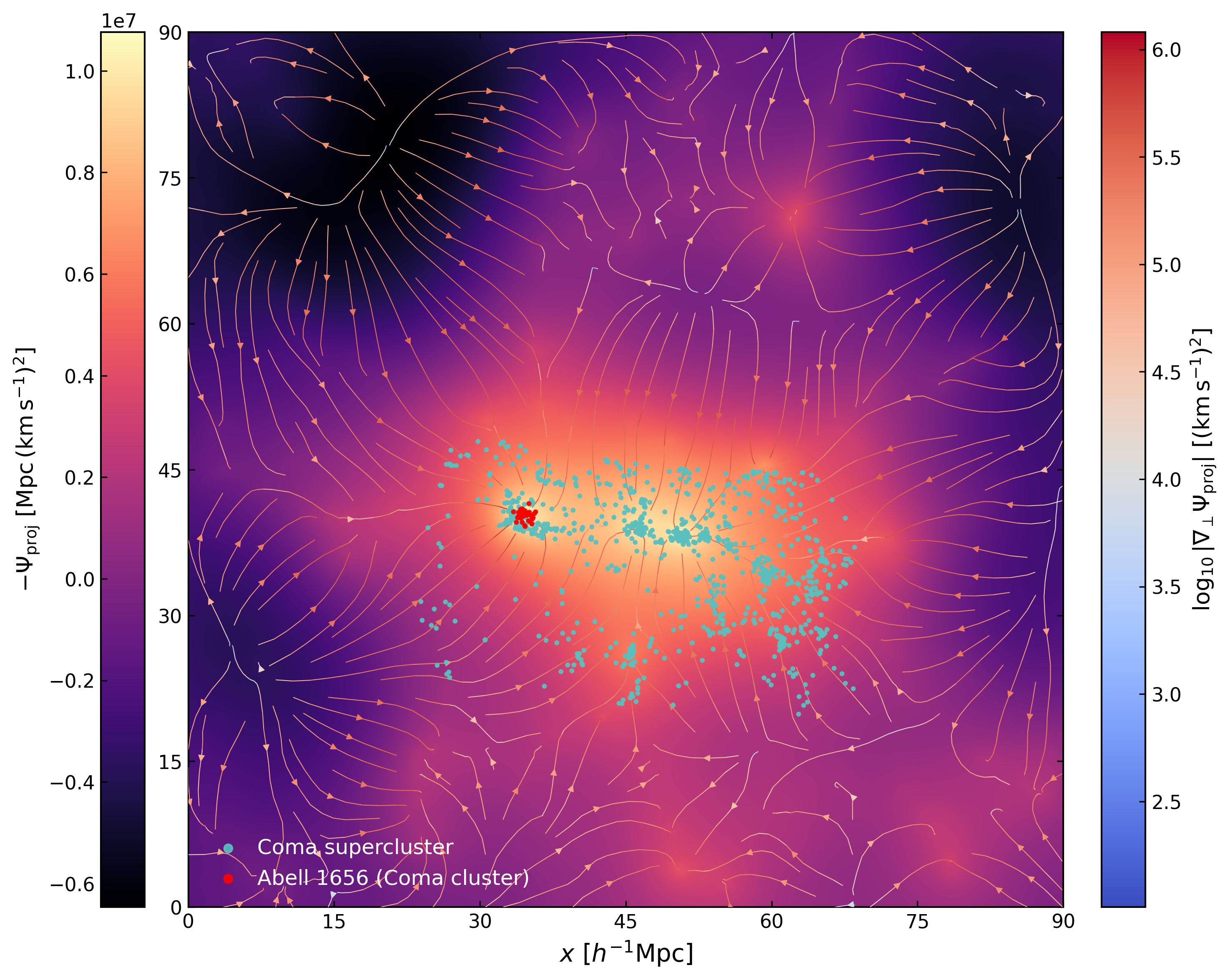} \caption{\textbf{Two-dimensional projected potential and the corresponding projected acceleration field derived from the reconstructed dark matter surface-density field.} The background color map shows the negative projected potential, $-\Psi_{\rm proj}$, with brighter regions corresponding to deeper potential minima. The left color bar gives the value of $-\Psi_{\rm proj}$. Streamlines trace the transverse projected acceleration field, $\mathbf{g}_{\perp,\mathrm{proj}}=-\nabla_\perp\Psi_{\rm proj}$, and converge toward regions of high projected mass concentration. Their color encodes $\log_{10}|\mathbf{g}_{\perp,\mathrm{proj}}|$, as indicated by the right color bar. Cyan and red points mark galaxies in the Coma supercluster and the Coma cluster, respectively. }
\label{fig4:potential}
\end{figure*}


\section{Datasets} \label{sec:data}
\subsection{CAMELS training set} \label{sec:camels}

The training set is drawn from the CAMELS simulations, which provide $(25\,h^{-1}\mathrm{Mpc})^3$ cosmological volumes from three simulation suites: Astrid, IllustrisTNG, and SIMBA \citep{nelson_2021,Bird_2022,Dav_2019}. Following \citet{ono_2024}, we adopt the Astrid suite at $z=0$, which was found to give the best reconstruction performance for this class of task. For training, we use the Astrid Latin-hypercube (LH) set, which consists of 1000 independent simulations spanning a range of cosmological and astrophysical parameters. This diversity is important because it allows the model to learn a more transferable relation between galaxies and dark matter, rather than features associated with a single simulation setup.

The model is designed for direct application to galaxy survey data. We therefore construct the input field from the galaxy catalog rather than from full stellar particle information. Specifically, we select galaxies with stellar masses above $10^9\,M_\odot$, matching the regime in which the SDSS sample is close to complete. The selected galaxies are then projected into two-dimensional stellar-mass surface density fields, with each galaxy treated as a point source and its stellar mass assigned to the corresponding pixel. In this way, the training input is defined in the same observationally motivated form as the field later used for the SDSS application.

The target dark matter density fields are taken from the CAMELS Multifield Dataset (CMD) \citep{Villae_2022}, which provides projected maps along the three coordinate axes with a thickness of $5\,h^{-1}\,\mathrm{Mpc}$ and a size of $256 \times 256$ pixels. To make the target more comparable to the observational setting, we stack five consecutive slices along the same axis into a single projected dark matter map with a total depth of $25\,h^{-1}\,\mathrm{Mpc}$. This choice serves two purposes: it matches the slab thickness adopted later for the SDSS analysis, and it reduces sensitivity to small-scale redshift-space distortions by averaging over peculiar-velocity-induced displacements along the line of sight. For each simulation box, we generate three projected maps, one along each coordinate axis, yielding a total of 3000 paired stellar-mass surface density and dark matter maps.



\subsection{IllustrisTNG300-1 validation set}

IllustrisTNG300-1 serves as an out-of-domain validation set. It is a cosmological magnetohydrodynamical simulation of a $(205\,h^{-1}\,\mathrm{Mpc})^3$ comoving volume, with a dark matter particle mass of $m_{\rm DM}=4.0\times10^7\,M_\odot/h$ and a baryonic mass resolution of $m_{\rm b}=7.6\times10^6\,M_\odot/h$ \citep{Springel_2017,Pill_2017, nelson_2021}. The simulation was run with the moving-mesh code \textsc{AREPO} under a $\Lambda$CDM cosmology, and included baryonic physics such as star formation, chemical enrichment, and stellar and AGN feedback. Compared with the CAMELS training volumes, IllustrisTNG300-1 provides a substantially larger and more statistically representative realization of large-scale structure and galaxy populations. It therefore offers a meaningful test of whether the trained model has learned a robust galaxy--dark matter mapping rather than one tied to the particulars of the training suite.

To ensure consistency with the training and observational setups, we construct projected dark matter maps by slicing the simulation box along one axis with the same slab thickness of $25\,h^{-1}\,\mathrm{Mpc}$. Each projected map has $2100\times2100$ pixels, matching the spatial resolution of the model. Halo catalogs are taken from the simulation group catalog, and the input stellar-mass surface density maps are built from the subhalo catalog. This validation set is therefore matched to the training and inference geometry while remaining genuinely independent in simulation volume, numerical implementation, and galaxy population. 


\subsection{SDSS galaxy sample}

The galaxy sample is taken from the MPA--JHU catalog \citep{Kauff_2003,Trem_2004,Brin_2004} based on SDSS DR8 \citep{York_2000,aiha_2011}. We restrict the analysis to the Northern Galactic Cap to ensure sufficient contiguous sky coverage for statistical analysis. Galaxy positions are converted from right ascension, declination, and redshift into Cartesian coordinates, with the line of sight defined as the $z$-axis. 

We further select a contiguous low-redshift SDSS region and extract a comoving subvolume of $(90\times90\times25)\,(h^{-1}\,\mathrm{Mpc})^3$ centered at $z=0.03$, with the $25\,h^{-1}\,\mathrm{Mpc}$ depth taken along the line of sight. This subvolume is chosen to lie well within the survey footprint and to avoid boundary truncation in the projected map. It is also close to complete above 
$10^9\,M_\odot$ in the adopted redshift range. Importantly, the selected region contains a broad range of environments, including rich clusters such as Coma and extended voids, making it a useful testbed for assessing reconstruction across multiple structure regimes.

The input SDSS field is constructed in the same way as the simulation input. Galaxies within the selected subvolume are projected along the line of sight to form a stellar-mass surface density map on the model grid. Each galaxy is treated as a point, and the stellar masses of galaxies falling into the same pixel are summed and divided by the pixel area.

\section{Method}\label{sec:model}

We formulate the problem as learning the distribution of the projected dark matter field $x_{\rm DM}$ conditioned on the observed galaxy stellar-mass field $x_{\rm gal}$, which is implemented with a conditional diffusion model \citep{King_2021,ono_2024}. 

The model is trained on paired simulation maps, with the projected galaxy stellar-mass map as input and the corresponding projected dark matter map as target. In both the training set and the SDSS application, we use galaxies with stellar masses above $10^9\,M_\odot$. Starting from the target map $x_{\rm DM}$, Gaussian noise is added at different diffusion times to produce a sequence of noisy maps,
\begin{equation}
z_t = \alpha_t x_{\rm DM} + \sigma_t \epsilon,
\qquad
\epsilon \sim \mathcal{N}(0,\mathbf{I}),
\end{equation}
where $z_t$ denotes the noisy dark matter field at diffusion time $t$, and $\alpha_t$ and $\sigma_t$ control the signal and noise amplitudes. The network is then trained to predict the injected noise from $z_t$ together with the corresponding stellar-mass map, thereby learning to iteratively remove noise and recover the underlying dark matter distribution.

The denoising network is a U-Net with residual blocks \citep{ronn_2015,He_2016}. The noisy field and the conditioning galaxy map are concatenated along the channel dimension at the network input in order to preserve small-scale spatial information. Group normalization \citep{wu_2018} is used throughout the network, and a self-attention block is inserted at the bottleneck layer to improve the recovery of long-range spatial correlations in the cosmic web \citep{vasw_2017,park_2018}. Before training, both the input and target maps are transformed into logarithmic space and normalized. 



Following \citet{ono_2024}, the total training loss is defined as
\begin{equation}
\mathcal{L}_{\rm total}
=
\mathcal{L}_{\rm diff}
+
\lambda_{\rm latent}\mathcal{L}_{\rm latent}
+
\lambda_{\rm rec}\mathcal{L}_{\rm rec},
\label{eq:loss_total}
\end{equation}
where $\mathcal{L}_{\rm diff}$ is the diffusion denoising loss, $\mathcal{L}_{\rm latent}$ is the Kullback--Leibler divergence between the distribution of the diffused field and a standard normal prior, and $\mathcal{L}_{\rm rec}$ is a reconstruction term evaluated near zero diffusion time. The model is optimized with AdamW \citep{losh_2017} using a learning rate of $10^{-5}$ and a batch size of 12.

At inference time, we start from a Gaussian noise field and iteratively apply the learned denoising process conditioned on the galaxy map to generate a dark matter map. For each input field, we generate 100 stochastic realizations by varying the random seed of the initial noise field, and adopt their pixel-wise mean as the fiducial reconstruction. The resulting realization-to-realization variation primarily reflects the seed-dependent stochasticity of the diffusion sampling. Averaging these realizations suppresses seed-level fluctuations while retaining the stable structures that are consistently supported by the galaxy data.

\section{Results} \label{sec:results}

We now turn to the two central questions of this work: whether the model generalizes beyond the CAMELS training domain, and whether it can yield a physically meaningful dark matter reconstruction when applied to a real galaxy survey. We first test the model on the independent IllustrisTNG300-1 simulation, where the true dark matter field is known and quantitative validation is possible at both the halo and pixel levels. We then apply the model to the SDSS galaxy sample and examine the reconstructed projected dark matter field, the inferred halo masses and HMF, and the derived gravitational potential across a nearby region spanning rich clusters, filaments, and voids.

\subsection{Cross-simulation generalization}

We first assess the ability of the model to generalize beyond the CAMELS training domain. Although the training set spans a broad range of cosmological and baryonic-physics parameters, a meaningful validation requires an independent simulation with different numerical implementation and a larger, more representative cosmological volume. We therefore apply the trained model to IllustrisTNG300-1, which provides an out-of-domain test of whether the reconstruction captures a robust relation between galaxies and dark matter, rather than features specific to the training simulations. We further find that the ensemble-averaged reconstruction generally shows improved agreement with the TNG300-1 truth relative to a typical individual stochastic realization, consistent with the suppression of realization-dependent fluctuations through averaging.


Figure~\ref{fig1:tng} summarizes the halo-level performance on the IllustrisTNG300-1 simulation. In the left panel, we compare the masses obtained by integrating the projected dark matter surface density within $R_{200}$ in the reconstructed map and in the ground-truth map. The two agree closely over most of the mass range with only a small mean offset. The inset further shows that the overall residual statistics are $\mu = 0.022$ dex and $\sigma = 0.173$ dex, indicating that the reconstruction recovers halo-scale projected masses with high fidelity when compared directly with the truth map.


The right panel of Figure~\ref{fig1:tng} provides a more observationally relevant comparison between the reconstructed projected-aperture masses with the catalog $M_{200}$ values in IllustrisTNG300-1. In this case the scatter is slightly larger, which is expected because projection can blend nearby systems within the same grid and the catalog mass definition is not identical to the projected mass integrated from the map. Even so, the agreement remains strong, with overall residual statistics of $\mu = 0.015$ dex and $\sigma = 0.197$ dex. Taken together, these results show that the model generalizes well to an independent simulation and robustly reproduces the statistical properties of the halo population without being tuned to a single simulation family.

We next examine the reconstruction at the pixel level using the full $2100\times2100$ grid of the IllustrisTNG300-1 map. Figure~\ref{fig5:pixel} shows that the reconstructed pixel values are broadly consistent with the true values in the surface-density regime most relevant for the SDSS application. Above $\log_{10}\Sigma_{\rm DM} \sim 11.5\,M_\odot\,h^{-1}(h^{-1}\,\mathrm{Mpc})^{-2}$, the mean residual in each bin stays below 0.2 dex and the standard deviation is typically around 0.3 dex. This suggests that the model not only traces halo-integrated quantities, but also retains meaningful spatially resolved information in the projected dark matter surface-density field. The four colored vertical lines mark the characteristic surface densities of the smallest halos above different mass thresholds. For the three higher thresholds, these values are estimated from the mass and radius of the smallest halo in each case. For halos of $10^{10.5}\,M_\odot\,h^{-1}$, however, the projected area is smaller than one map pixel, whereas halos of $10^{10.6}\,M_\odot\,h^{-1}$ occupy roughly one pixel. We therefore adopt $10^{10.6}\,h^{-1}\,M_\odot$ as an approximate geometric reference scale for the spatial sampling of the projected map.

At lower surface densities, the reconstruction becomes less accurate and shows a clear degradation in both bias and scatter. This behavior is physically expected. These regions are associated with lower-mass galaxies and weaker structures, for which the available information is more limited given the stellar-mass cut of the input sample and the finite pixel resolution. In other words, the model is most reliable for the intermediate- and high-density structures that dominate the observable nearby cosmic web, while finer and fainter structures remain harder to recover.

Overall, the IllustrisTNG300-1 validation provides support for three main points. First, halo masses inferred from the reconstructed projected-aperture measurements agree well with the corresponding true values in an independent simulation. Second, the reconstructed maps retain useful statistical information on pixel-level dark matter surface densities, with a typical scatter of approximately 0.3 dex over the density range most relevant to the observational application. Third, halos with masses of approximately $10^{10.6}\,h^{-1}\,M_\odot$ have characteristic projected areas that are larger than one map-pixel area. These results indicate that the model has learned a transferable galaxy--dark matter mapping and motivate its application to the SDSS data in the next subsection.


\subsection{From SDSS Galaxies to the Projected Dark Matter Field}\label{subsec2:sdss}

\subsubsection{Reconstruction of the projected dark matter field}

Having validated the model on independent simulations, we next apply it to the SDSS galaxy sample. Figure~\ref{fig2:dmimage} shows the reconstructed projected dark matter surface density field in the selected volume at $z\sim0.03$, with a pixel size of $0.097\,h^{-1}\,\mathrm{Mpc}$. The map is obtained by averaging over 100 stochastic realizations generated from different initial noise seeds, and traces structure across the full $90 \times 90~(h^{-1}\mathrm{Mpc})^2$ region. The reconstruction reveals a rich range of environments, including compact overdense peaks associated with massive systems, extended filamentary structures connecting them, and broad underdense void regions. {These features should be interpreted as projected structures, since the reconstruction integrates the matter distribution over a $25\,h^{-1}\,\mathrm{Mpc}$ line-of-sight slab and may therefore be affected by line-of-sight superposition and blending.}

The reconstructed field is spatially associated with the observed SDSS galaxies marked by black circles, in line with the standard picture that galaxies reside within dark matter halos and trace the surrounding large-scale structure \citep{Wech_2018}. We also overlay galaxy groups from the Zhao catalog as red circles \citep{Zhao_2025}; this catalog provides updated halo mass estimates for SDSS groups relative to the original Yang group catalogs \citep{Yang_2005,Yang_2007}. Many of them coincide with enhanced dark matter concentrations, indicating that the model captures the dominant halo-scale mass concentrations in the nearby Universe. At the same time, the map contains substantially more structure than a set of isolated peaks: many dense regions are embedded in broader overdense environments and are linked by filamentary bridges, suggesting that the reconstruction traces the surrounding cosmic web rather than only the locations of individual massive halos.


The zoom-in panels in Figure~\ref{fig2:dmimage} further illustrate the local environments of these dense regions. In particular, massive systems are often surrounded by satellite-rich environments with elevated local dark matter density, consistent with the expectation that halos grow and evolve within anisotropic large-scale structure. This multi-scale morphology is one of the key outcomes of the reconstruction, showing that the inferred field contains physically meaningful spatial information from halo scales up to the surrounding web.

As an additional check, we identify member galaxies of Abell clusters in the SDSS sample \citep{Abell_1989,Ander_1991}; further details are given in Appendix~\ref{sec:abell}. These clusters are preferentially located in dense nodes and filament intersections, and some appear to be connected by broader overdense structures rather than being fully isolated. The Coma cluster provides the clearest example: even though only a subset of its member galaxies contributes to the input map, the reconstruction still recovers a strong and extended dark matter overdensity at its position. While the present resolution does not recover detailed internal substructure, it robustly captures the large-scale mass concentrations and the network-like environments of rich cluster systems.


\subsubsection{Halo mass statistics}

We next perform a halo-level consistency check of the SDSS reconstruction using the SDSS group catalog. For each system in the Zhao catalog, we measure the mass in the reconstructed map by integrating the projected dark matter density within $R_{200}$, adopting the catalog radius for direct comparison. The resulting projected-aperture masses provide a consistency test against the catalog halo masses, as shown in the left panel of Figure~\ref{fig3:halo}. The reconstructed masses agree closely with the Zhao catalog values over most of the mass range. The overall scatter is approximately $0.25$ dex and shows little dependence on halo mass. It is broadly consistent with the combination of reconstruction and the intrinsic uncertainty of group-catalog-based halo mass estimates.


At the same time, the comparison reveals a systematic relative trend. At the low-mass end, the reconstruction-based masses tend to be somewhat higher than the Zhao catalog values, whereas at the high-mass end they become mildly lower. A plausible explanation is that multiple nearby systems can overlap in projection and enhance the local surface density around lower-mass halos, while the most massive halos are more affected by finite spatial resolution and profile smoothing near $R_{200}$. The limited number of very massive systems in the selected volume may also contribute to the statistical uncertainty at the high-mass end. Since no large-sample, independent measurement of the true halo mass is available for individual SDSS systems, this trend may reflect a residual relative-calibration issue between the reconstruction-based and group-catalog-based mass scales.

The right panel of Figure~\ref{fig3:halo} compares the mass functions derived from the reconstruction with those from the Zhao catalog and the L-GALAXIES mock catalog \citep{Henr_2019} over the same redshift range. Built from a large-volume semi-analytic model, the mock catalog provides a statistically well-sampled galaxy and halo population with substantially reduced cosmic variance. The mass function provides a complementary statistical test because it depends on the full distribution of the inferred masses rather than on one-to-one object matching alone. Although no explicit mass-function prior is imposed during inference, the reconstructed mass function broadly follows the expectation from the mock catalog over a wide mass range.

The SDSS reconstruction covers a single local volume of
$(90\times90\times25)\,(h^{-1}\,\mathrm{Mpc})^3$, whereas the
L-GALAXIES catalog covers the full
$(685\,h^{-1}\,\mathrm{Mpc})^3$ simulation volume. Cosmic variance
may therefore affect both the normalization and shape of the inferred
halo mass function, with its impact potentially varying with mass because
halos of different masses have different abundances and clustering
strengths. The error bars in Figure~\ref{fig3:halo} include
statistical uncertainties but do not include an explicit
cosmic-variance contribution. Some of the differences among the reconstructed, catalog-based, and mock halo mass functions may therefore arise from large-scale density fluctuations within the selected SDSS volume.

At the same time, their mass-dependent differences are also consistent with the trend seen in the left panel. In particular, the relative excess at intermediate masses and deficit at the highest masses may reflect both systematic offsets between the reconstruction-based and catalog-based mass estimates and large-scale density fluctuations within the SDSS volume. The key point for the present work is that the reconstruction yields a physically plausible halo population without imposing the halo mass function during inference. The remaining discrepancies may reflect a combination of cosmic variance and residual relative-calibration differences between the reconstruction-based and catalog-based mass estimates, both of which will be investigated in future work.

\subsubsection{Gravitational potential and large-scale environment}

The reconstructed dark matter surface-density field also allows us to examine the projected large-scale gravitational environment. We solve the two-dimensional projected Poisson equation,
\begin{equation}
\nabla_\perp^2\Psi_{\rm proj}
=
4\pi G \delta\Sigma_{\rm DM},
\end{equation}
where $\delta\Sigma_{\rm DM}=\Sigma_{\rm DM}-\langle\Sigma_{\rm DM}\rangle$, using a fast Fourier transform method \citep{cool_1965}, with the $k=0$ mode set to zero to remove the uniform component and fix the potential zero point. The mean-subtracted map is used without additional Gaussian smoothing. We then derive the projected acceleration field as
$\mathbf{g}_{\perp,\mathrm{proj}}
=-\nabla_\perp\Psi_{\mathrm{proj}}$, which traces the direction and relative magnitude of the potential gradient in the projected plane.
Figure~\ref{fig4:potential} shows the projected potential and the corresponding projected acceleration field across the selected SDSS region. We also mark the positions of galaxies belonging to the Coma supercluster, a nearby large-scale structure that includes the Coma cluster, Abell 1367, and the surrounding filamentary environments \citep{greg_1978,maha_2010,maha_2018}.

The background color map represents $-\Psi_{\rm proj}$, with brighter regions indicating deeper potential wells. These regions spatially coincide with the strongest dark matter overdensities, as expected if the reconstruction correctly traces the dominant mass distribution. The potential map is much smoother than the density map, which is physically expected: the gravitational potential is a non-local quantity that reflects the integrated mass distribution over a broad region and therefore suppresses small-scale contrast while emphasizing the coherent large-scale structure. The projected potential field thus provides a useful way to visualize the large-scale mass environment encoded in the reconstruction.


More importantly, the broader Coma supercluster lies within one of the deepest regions of the projected potential, where the streamlines of the inferred projected acceleration field converge toward a prominent local minimum near the Coma cluster. This result is significant because Coma is not only recovered as a local overdensity peak in the density field, but also as a major center of gravitational influence in the larger-scale environment. In other words, the reconstruction preserves not only local halo-scale peaks, but also the coherent gravitational structure of the surrounding cosmic web. This is one of the central scientific points of the present work: the inferred field contains enough large-scale information to move beyond galaxy-overdensity-based descriptions of environment and toward a more physical characterization based on projected dark matter and halo environment.

\section{Summary and Discussion}\label{sec:summary}

In this work, we have presented a conditional variational diffusion model for reconstructing the projected dark matter field from an observable projected galaxy stellar-mass density field, and applied it to a nearby SDSS survey region. The model is trained on the CAMELS LH set and validated on the independent IllustrisTNG300-1 simulation using galaxies with $\mathrm{M_\star \ge 10^9 \, M_\odot}$. This design makes the method directly applicable to galaxy survey data, without relying on simulation-only information such as the full stellar particle distribution or being tied to a particular parameter choice. It therefore provides a practical route for diffusion-based dark matter reconstruction in real observations.

The main results and their implications are summarized as follows:

\begin{itemize}

    \item \textbf{The model generalizes well beyond the training set.}
     In validation on the independent IllustrisTNG300-1 simulation, the reconstructed masses closely follow the reference values, with a scatter below 0.2 dex at the halo level. At the spatially resolved pixel level, the reconstruction remains reliable in the surface-density regime most relevant to the SDSS application, with mean residuals below 0.2 dex and a typical scatter of about 0.3 dex. With a pixel size of $0.097\,h^{-1}\,\mathrm{Mpc}$, halos below approximately $10^{10.6}\,h^{-1}\,M_\odot$ have characteristic projected areas smaller than one map pixel, although their mass still contributes to the surface-density field. Taken together, these tests indicate that the model learns a robust galaxy--dark matter mapping rather than features tied to a single simulation realization.

    \item \textbf{The method captures the dominant multi-scale structures in a real survey region.}
    In the selected contiguous SDSS field, the reconstruction covers a $90\times90\,(h^{-1}\,\mathrm{Mpc})^2$ region with a pixel size of $0.097\,h^{-1}\,\mathrm{Mpc}$, and reveals cluster-scale overdensities, filamentary structures, and extended voids. The zoom-in views in Figure~\ref{fig2:dmimage} show that high-density peaks are not isolated but are embedded in broader overdense environments. The Abell systems examined in Appendix~\ref{sec:abell} further support this picture: they tend to lie in dense nodes and filament intersections, and some appear as part of a connected network of overdense structures rather than as isolated peaks. In this sense, the reconstruction captures not only individual halo-scale concentrations, but also the surrounding cosmic web in which they are embedded, subject to the projection and line-of-sight blending inherent in the two-dimensional reconstruction.

    \item \textbf{The reconstructed field preserves physically meaningful information beyond local overdensity peaks.}
    As consistency checks, we compare the halo masses inferred from the reconstructed projected-aperture measurements with the corresponding values in the SDSS Zhao group catalog, and the resulting halo mass function with the expectation from the L-GALAXIES mock catalog. The inferred halo masses are broadly consistent with the Zhao catalog, while the resulting HMF follows the overall mock expectation despite no explicit HMF prior being imposed during inference. This latter comparison is primarily limited by cosmic variance within the selected SDSS volume. In addition, the Zhao-catalog comparison shows a mass-dependent relative offset, with the inferred halo masses being mildly higher at the low-mass end and lower at the high-mass end. This offset is qualitatively consistent with the differences between the reconstructed and catalog-based halo mass functions, and should be interpreted as a relative-calibration caveat given the lack of large-sample, independent halo-mass measurements for individual SDSS systems.

    \item \textbf{The derived projected gravitational potential highlights the non-local, large-scale gravitational coherence of the reconstruction.}
    By solving the two-dimensional Poisson equation, we find that the deepest potential regions coincide with the strongest overdensities. In particular, the Coma supercluster resides in the deepest potential regions and near the convergence of projected acceleration streamlines, with the Coma cluster occupying one of its deepest minima. This shows that the reconstruction preserves not only local halo-scale peaks, but also coherent large-scale gravitational environments. This is important because the projected density map mainly emphasizes local contrast, whereas the potential field highlights the non-local, integrated mass distribution. The ability to recover this larger-scale gravitational coherence is one of the most important outcomes of the present work.

\end{itemize}

The broader significance of this study is that it opens a path toward a new observationally driven description of galaxy environment. Traditionally, environmental studies of galaxy formation and evolution have been based primarily on galaxy number density or related overdensity estimators. Our reconstruction makes it possible to move toward halo-mass- and dark-matter-environment-based descriptions using real survey data. In that sense, the importance of this work is not only that it reconstructs a projected dark matter map from the SDSS survey, but also that it begins to connect observational studies of galaxy evolution more directly to the underlying mass field that drives structure formation.

Several limitations should be kept in mind. First, the present reconstruction is two-dimensional and based on a slab of thickness $25\,h^{-1}\,\mathrm{Mpc}$, so projection effects remain unavoidable. Line-of-sight projection and overlap may influence the projected shapes of structures and their apparent connections, and may also contribute additional uncertainty to projected-aperture mass estimates. Second, redshift-space distortion is not explicitly modeled. The adopted slab thickness helps reduce sensitivity to small-scale line-of-sight displacements, but it also blends overlapping structures. Third, the input is restricted to galaxies with $M_\star \ge 10^9\,M_\odot$, chosen to ensure observational completeness. As a result, the method is expected to recover intermediate- and large-scale structures more robustly than the faintest and smallest-scale structures, for which information from lower-mass galaxies is missing. These limitations likely contribute to the reduced fidelity of the reconstruction in the low-density regime and to the limited recovery of substructure.

These considerations also indicate clear directions for future work. A fully three-dimensional diffusion model with redshift-space effects explicitly included would reduce projection-related ambiguities and provide a more realistic treatment of galaxy survey data. Improving the spatial resolution and extending the input to lower-mass galaxies would also help recover smaller halos and finer structures more faithfully. Applied to wider and deeper surveys such as DESI, LSST and CSST, such models could enable statistical studies of the galaxy--dark matter connection across a much broader range of environments and cosmic epochs.

More generally, this work demonstrates a practical application of diffusion-based dark matter reconstruction to real galaxy survey data. It provides one of the first demonstrations that a diffusion model, trained on simulations but driven by directly observable galaxy inputs, can recover a spatially resolved projected dark matter field over nearly three orders of magnitude in scale, from halo-sized peaks to the surrounding large-scale web, as well as the gravitational potential. This creates a new observational route for studying the physics of galaxy evolution in the context of the underlying mass distribution, and offers a promising foundation for future work on nonlinear structure formation and cosmological inference.

\section*{Data Availability}

The reconstructed SDSS projected dark matter map and the galaxy sample used to construct the input stellar-mass-density field are publicly available on Zenodo (\href{https://doi.org/10.5281/zenodo.21931967}{doi:10.5281/zenodo.21931967}).

\begin{acknowledgments}

We thank the referee for the careful reading of the manuscript and for the constructive comments that helped improve the clarity of this work. Y.P. and K.Z. acknowledge support from the National SKA Program of China under grant No. 2025SKA0150102 and from the National Natural Science Foundation of China (NSFC) under grant Nos. 12125301, 12192220, and 12192222. Y.P. also acknowledges support from the New Cornerstone Science Foundation through the XPLORER PRIZE. J.D. acknowledges the support of National Science Foundation of China (NSFC) grant Nos. 12303010. This work is extensively supported by the High-performance Computing Platform of Peking University, China. 
\end{acknowledgments}


%


\software{
          Transformers \citep{transfomer_2019};
          astropy~(\citealp{astropy_2013,astropy_2018,astropy_2022});
          NumPy \citep{numpy_2020};
          PyTorch \citep{pytorch_2019};
          Matplotlib \citep{matplot_2007}
          }


\appendix

\setcounter{figure}{0}

\renewcommand{\thefigure}{A\arabic{figure}}
\renewcommand{\theHfigure}{A\arabic{figure}}
\begin{figure*}[h]
\centering
\includegraphics[width=\textwidth]{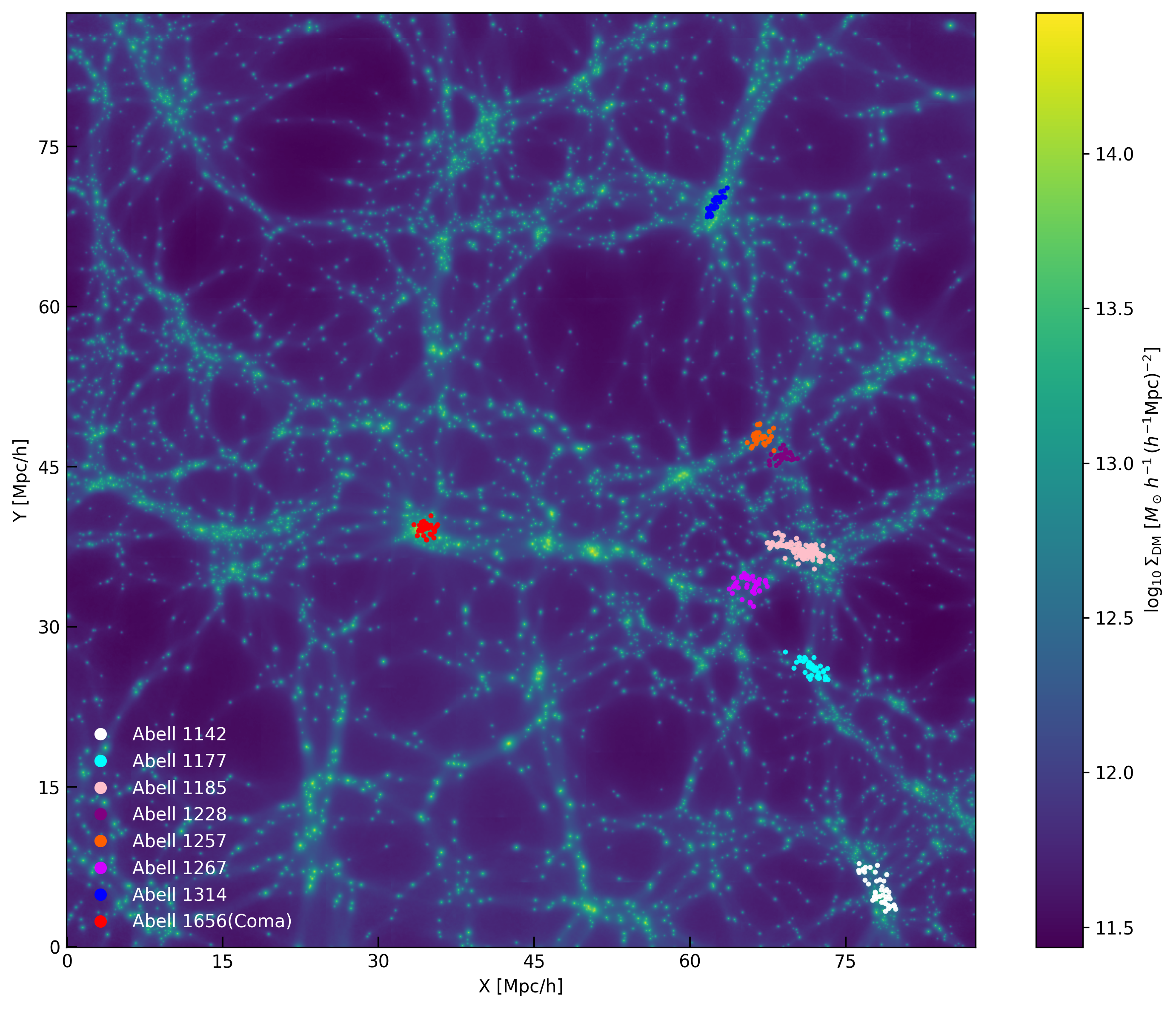}
\caption{\textbf{\boldmath Reconstructed projected dark matter density field in the SDSS region with Abell cluster members overlaid.} The background colour scale shows the logarithmic projected dark matter surface density. Colored dots mark galaxies identified as members of different Abell clusters, and red dots highlight the galaxies residing in Coma cluster (Abell 1656). The clusters tend to lie at dense nodes and filament intersections in the reconstructed field, and their names are labeled in the lower left corner. }\label{fig6:cluster}
\end{figure*}

\begin{figure*}[h]
\centering
\includegraphics[width=\textwidth]{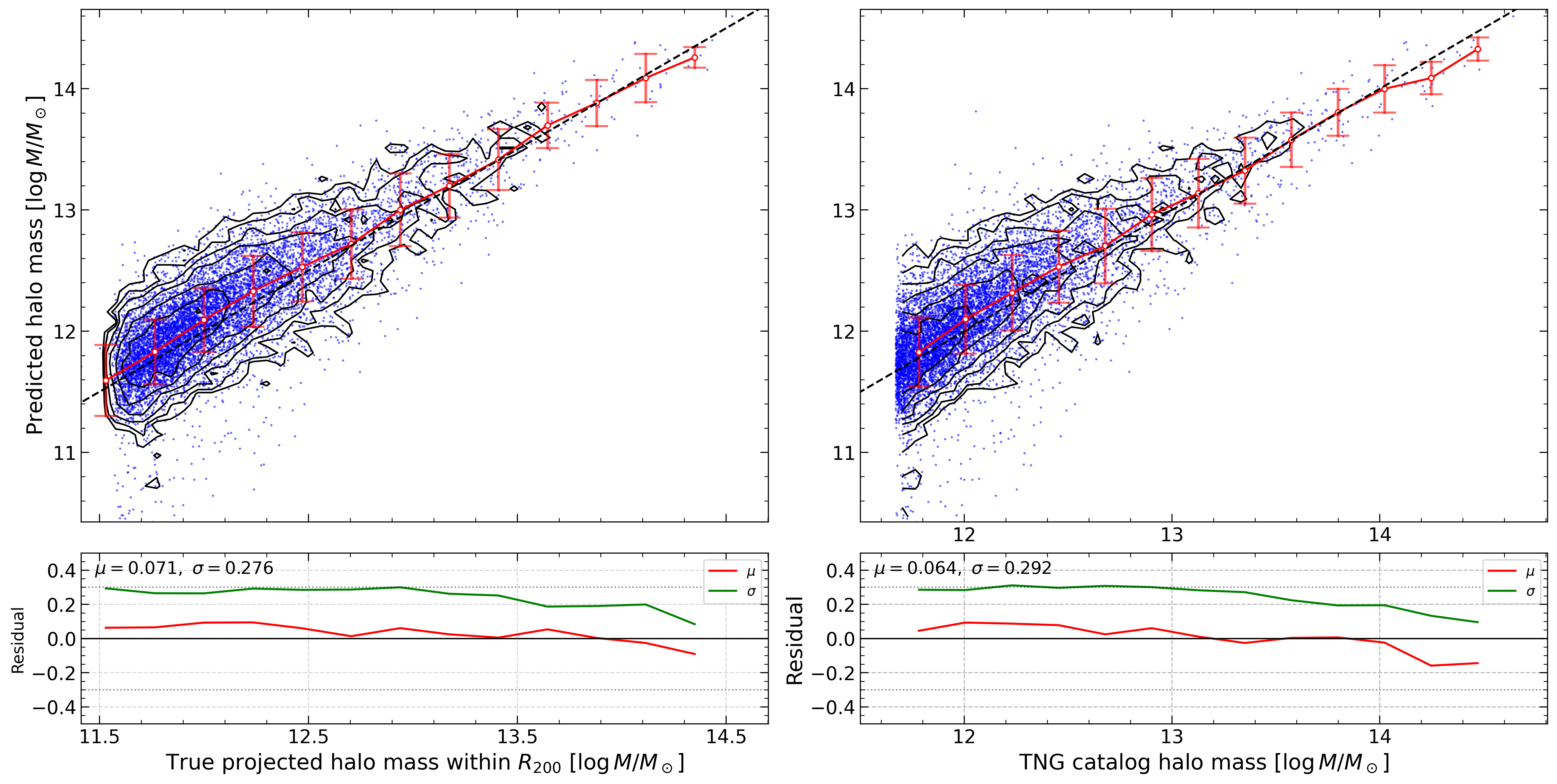}
\caption{\textbf{Validation of the projected-aperture-based halo mass estimates after including fiber collisions and stellar-mass uncertainties in IllustrisTNG300-1.}
\textit{Left:} Comparison between the projected-aperture masses integrated within $R_{200}$ in the ground-truth and reconstructed projected matter maps. Blue points denote individual dark matter halos, and black contours show their number-density distribution. The red line shows the mean reconstructed projected-aperture mass in bins of the true projected-aperture mass, with error bars indicating the standard deviation. The black dashed line denotes the one-to-one relation. In the inset, the red and green lines show the mean ($\mu$) and standard deviation ($\sigma$) of the residuals in each true projected-aperture-mass bin, respectively, while the gray dashed lines mark $\pm0.3$ dex.
\textit{Right:} Comparison between the reconstructed projected-aperture mass within $R_{200}$ and the catalog mass $M_{200}$. Visual conventions follow those used in the left panel.}\label{fig7:obs}
\end{figure*}

\section{Abell cluster members} \label{sec:abell}

As an observationally intuitive check on the reconstruction, we examine regions containing known Abell clusters in the selected SDSS volume \citep{Abell_1989,Ander_1991}. If the reconstructed map captures the large-scale dark matter environment in a physically meaningful way, rich clusters should preferentially lie in dense nodes and filament intersections of the projected field. Figure~\ref{fig6:cluster} confirms this expectation. It also illustrates that, although the present reconstruction does not resolve detailed internal cluster substructure, it robustly captures the extended dark matter overdensities associated with rich nearby systems such as Coma.

\section{Additional validation with SDSS observational effects} \label{sec:obs}

We have performed an additional test using IllustrisTNG300-1 to examine the effects of fiber collisions and stellar-mass uncertainties in SDSS observations. To model the impact of fiber collisions, we adopt a conservative procedure in which, for galaxy pairs separated by less than $55^{\prime\prime}$, we randomly remove one galaxy from each close pair until no such pairs remain. This treatment is conservative because it neglects the partial recovery of close pairs in regions covered by overlapping SDSS spectroscopic plates. We then perturb the stellar masses with Gaussian errors of 0.1 dex and reapply the same stellar-mass selection used for the reconstruction. Figure~\ref{fig7:obs} compares the resulting projected-aperture masses with the corresponding true masses in the simulation, using the same analysis procedure as in the main validation. We find no significant systematic offset after including these observational effects, although the scatter increases modestly, as expected from the added stellar-mass uncertainty.





\bibliography{sample701}{}
\bibliographystyle{aasjournalv7}



\end{document}